\documentclass[journal]{IEEEtran}
\usepackage{mypackages}

\begin{document}
%
\title{Performance Evaluation of Scheduling in 5G-mmWave Networks under Human Blockage}
%
%
%

\author{Fadhil~Firyaguna,
        Andrea~Bonfante,
        Jacek~Kibi{\l}da
        and~Nicola~Marchetti
%
\thanks{The authors are with CONNECT Centre, Trinity College Dublin, Ireland. E-mail: \{firyaguf, bonfanta, kibildj, nicola.marchetti\}@tcd.ie.}
\thanks{This publication has emanated from the research conducted within the scope of \textit{NEMO (Enabling Cellular Networks to Exploit Millimeter-wave Opportunities)} project financially supported by the Science Foundation Ireland (SFI) under Grant No. 14/US/I3110 and with partial support of the European Regional Development Fund under Grant No. 13/RC/2077.}
}

\maketitle

\begin{abstract}
The millimetre-wave spectrum provisions enormous enhancement to the achievable data rate of 5G networks. 
However, human blockages affecting the millimetre-wave signal can severely degrade the performance if proper resource allocation is not considered.
In this paper, we assess how conventional schedulers, such as the Proportional Fair scheduler, react to the presence of blockage.
Our results show that the resource allocation may disfavour users suffering from blockage, leading to low data rate for those users.
To circumvent this problem, we show that the data rate of those users can be improved by using a scheduler adapted to react to upcoming blockage events.
The adapted scheduler aims at proactively allocating the resources before a blockage happens, mitigating losses.
Such adaptation is motivated by recent progress in blockage prediction for millimetre-wave signals in a dynamic human blockage scenario.
Our extensive simulations indicate gains in the 1st percentile rate and fairness with respect to Proportional Fair scheduler when blockage conditions are severe.
\end{abstract}


%
\IEEEpeerreviewmaketitle

\section{Introduction}
\label{sec:introduction}
The \ac{mmWave} spectrum is key to enhance the capacity of the fifth-generation (5G) of cellular networks and support multi-Gbps applications.
However, transmission in \ac{mmWave} frequencies may be severely affected by obstructions, such as human bodies.
The blockage caused by a human body may add as much as \unit[40]{dB} of attenuation to the \ac{mmWave} signal \cite{maccartney2017rapid}, which can lead to radio link failures.
Human blockages can be frequent in crowded places such as transportation hubs and sport arenas.
This frequent, and potentially abrupt, disruption to the signal quality may impact the effectiveness of radio resource scheduling algorithms.

Blockage-aware scheduling algorithms have been adopted in the context of \acp{mmWave} for relay-assisted networks, where the blockage problem is handled through allocating the transmissions over multi-hop relay paths and thus preventing possible disruptions and enabling seamless connectivity \cite{lella2019blockage}.
However, there is a lack of studies in the context of networks where relay nodes or other \acp{AP} are not available for re-connection in the event of a blockage. In this context, a \ac{UE} suffering from frequent blockage can have its transmission rate severely reduced, increasing the discrepancy to other \acp{UE} operating in favourable channel conditions, as we will also show in our numerical evaluation.


In this paper, we evaluate the performance of resource allocation algorithms, based on the conventional \ac{PF} scheduler, in the presence of \ac{mmWave} signal blockages.
We consider that the resource allocation is performed to multiple users, each of which may experience abrupt disruptions to signal quality caused by the human body blockage.
Since the \ac{PF} scheduler prioritises the allocation based on the achievable transmission rate, \acp{UE} suffering from blockage may have the feasible rate degraded and thus be given progressively lower priority, disfavouring the assignment to those \acp{UE}.
Hence, the challenge is to adapt the scheduler in such way that the allocation considers these signal disruptions and possible rate degradation in the following moments, and thus produces more fair assignment to \acp{UE} suffering from blockage.

\begin{figure}
    \centering
    \includegraphics[width=\linewidth,trim={0.5cm 12cm 7.5cm 11cm},clip]{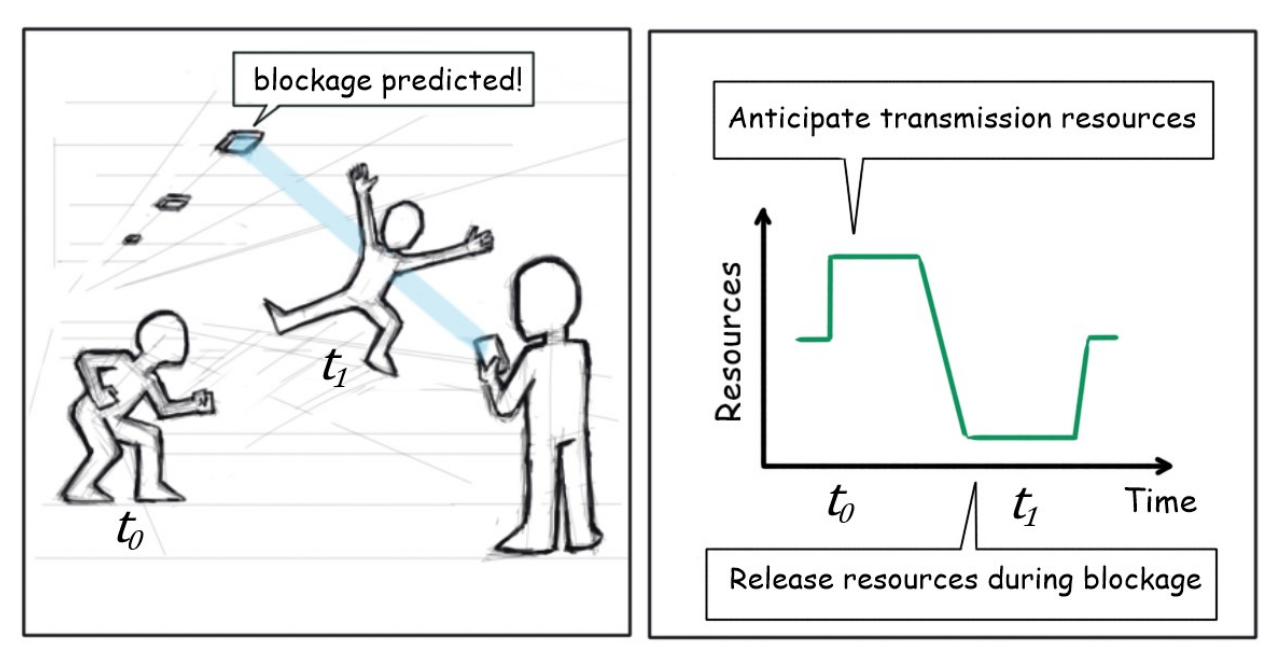}
    \caption{Preemptive resource allocation for users that might be blocked.}
    \label{fig:blockage_resource_anticipation}
\end{figure}

With the development of blockage prediction mechanisms, a scheduler may effectively mitigate the blockage effects, as we will show in this work.
Blockage prediction in \ac{mmWave} communications can be performed by: (i) inferring the \ac{mmWave} channel from the sub-\unit[6]{GHz} channel observation \cite{alrabeiah2019deep,ali2019early},
(ii) observing the past \ac{mmWave} beamforming vectors \cite{alkhateeb2018machine,liu2018interference} or (iii) using camera images to track the movement of potential obstructions \cite{nishio2019proactive,alrabeiah2019millimeter}.
Those methods can produce blockage indicators that will help the system predict the channel status.
The scheduler can use these methods to anticipate transmission to a user that might be blocked and preemptively allocate resources, as illustrated in Fig. \ref{fig:blockage_resource_anticipation}.



\subsection{Related Works}

Prediction-based schedulers have been studied in the context of previous generations of wireless systems \cite{bang2008channel,hajipour2010proportional,schmidt2011prediction,margolies2016exploiting}.
These schedulers rely on predicted and measured information that includes fading channel gains, mobility pattern (predicted route of a user through a region) and radio coverage map (mapping between measured signal strengths and geographical locations).
The prediction of the fading channel in the upcoming transmission intervals allowed these schedulers to estimate the achievable rates with increased precision, improving the network performance under the effects of rapid fluctuations of the received signal strength due to fast-fading at sub-\unit[6]{GHz} frequencies.
However, as the accuracy of fading prediction can degrade rapidly with the prediction horizon (few milliseconds, depending on the mobility scenario) \cite{schmidt2011prediction}, the schedulers proposed in \cite{bang2008channel,hajipour2010proportional,schmidt2011prediction,shen2019proactive} have to update their prediction and re-optimise scheduling every millisecond.
When considering mobility, the scheduler in \cite{margolies2016exploiting} uses the prediction of the mobility pattern together with the radio coverage map to estimate the achievable rate and allocate resources accordingly. It was shown that this approach can significantly improve network performance for urban mobility scenarios and slow-fading channels. 
However, none of these works, e.g., \cite{bang2008channel,hajipour2010proportional,schmidt2011prediction,margolies2016exploiting}, account for the effect of the \ac{mmWave} channel propagation characteristics, including the blockage effect.

For 5G wireless systems, most of the works have been focusing on developing schedulers to improve the fairness in heterogeneous networks \cite{noliya2020performance} and the cell edge throughput \cite{deniz2018performance}. 
Still, they do not consider the blockage effects on the channel and on the scheduler design.
Therefore, it is not clear how those schedulers will react to blockages.
In such conditions, the channel quality of a user can be more affected than, or as much as, the channel of cell edge users in 5G-\ac{mmWave} networks.

On the other hand, blockage-aware schedulers have been adopted for relay-assisted networks (WPAN, WLAN, 5G wireless backhaul) \cite{wang2019relay,lella2019blockage,niu2019relay,liu2019blockage,chen2018multipath,chang2018efficient,vu2018path}, where the scheduling algorithm can allocate a transmission to a relay AP when the \ac{UE} link to the serving \ac{AP} becomes blocked.
Such scheduling algorithms make their decision after a blockage event (reactive scheduling) and may have to re-calculate the decision every time a blockage occurs, potentially increasing overhead as blockage frequency increases.


\subsection{Contributions}

In contrast to the reactive methods proposed in  \cite{wang2019relay,lella2019blockage,niu2019relay,liu2019blockage,chen2018multipath,chang2018efficient,vu2018path}, our work considers that the scheduler could instead have a proactive approach with the aid of prediction mechanisms.
Since human blockage represents a significant challenge for the 5G systems design, new prediction methods have been recently studied in \cite{ali2019early}, where one proposes the utilisation of sub-\unit[6]{GHz} signal to provide an early warning of blockage in the \ac{mmWave} signal, and in \cite{alkhateeb2018machine,alrabeiah2019deep,alrabeiah2019millimeter,zarifneshat2019learning,nishio2019proactive} where one proposes to use camera images to construct a prediction model from a dataset of sequential images involving the geometry and mobility of obstacles and labelled with the received power. 
Using such methods, blockages can be forecasted up to tens of milliseconds ahead, giving the system sufficient time to adapt.
These positive blockage prediction results motivate us to revisit scheduler design for \ac{mmWave} networks.
In our work, we assume there exists a prediction mechanism, such as the one proposed in  \cite{nishio2019proactive} that allows the system to predict the received power in a given time window and to estimate future data rates based on that prediction.
The early-warning blockage indicator allows the scheduler to have sufficient time to allocate resources in order to mitigate the losses in performance caused by the transition from \ac{LOS} to blocked state.
With this system, we can implement a scheduler that allocates resources in future transmission intervals according to the estimated data rates and test its performance against the conventional schedulers.

Secondly, as the main objective of the allocation techniques proposed in \cite{xu2019adaptive,ali2019early,alkhateeb2018machine,zarifneshat2019learning,liu2018interference,nishio2019proactive,alrabeiah2019millimeter,moltafet2018joint,gerasimenko2019capacity,polese2018distributed} is to select a different path to avoid blockage, 
in our work we study an alternative way to mitigate blockage effects when there is no other cell to hand-off to.
Indeed, our work aims to evaluate how the conventional resource allocation methods behave under blockage when there is no spatial macro-diversity (single \ac{AP}), and to study how they can be adapted to cope better with the blockage effects.

In summary, the main contributions of this work account for:
\begin{itemize}
	\item We show that users suffering from blockage are disfavoured in the     resource allocation because conventional schedulers do not react quickly enough in case of blockage.
    This result shows that conventional schedulers, to be applied in 5G and beyond networks, should be adapted to cope with the impairments of the \ac{mmWave} propagation.
	\item We show that an adaptation of the \ac{PF} scheduler can be made by changing the priority metric to consider the blockage prediction.
    We refer to this adapted scheduler as the \ac{BA-PF} scheduler.
    We also show that using \ac{BA-PF} it is possible to increase resource allocation before the channel quality drops and reduce the allocation thereafter.
\end{itemize}

The rest of the paper is organised as follows: we describe our system model in Section \ref{sec:systemmodel}, we evaluate the user data rate performance, comparing our \ac{BA-PF} scheduler to conventional schedulers in Section \ref{sec:numericalresults} and we summarise our findings in Section \ref{sec:conclusion}.

\section{System Model}
\label{sec:systemmodel}
\subsection{Frame Structure}
We study the downlink transmission of a single ceiling-mounted indoor mmWave \ac{AP} which transmits 
to $n_U$ \acp{UE} using \ac{OFDMA} during \unit[$T$]{ms}.
Data transmission occurs in time-slots with duration $\Delta_t$ and in a bandwidth $b$ divided in $n_\textrm{SC}$ sub-carriers.
At the \ac{MAC} layer, we assume a scheduler assigns all sub-carriers at a time-slot $t$ to a \ac{UE} having index $u \in \{ 1, \dots, n_U \}$.
This assignment is represented by a vector $\mathbf{X}$, with size $T/\Delta t$ elements, defined for each \ac{UE}.
When the time-slot $t$ is assigned to \ac{UE} $u$, the scheduler sets  $x_u(t) = 1$. Otherwise, the scheduler sets $x_u(t) = 0$.

\subsection{Channel Model}
\label{sec:channel_model}
\begin{figure}
	\centering
	\includegraphics[width=.9\linewidth]{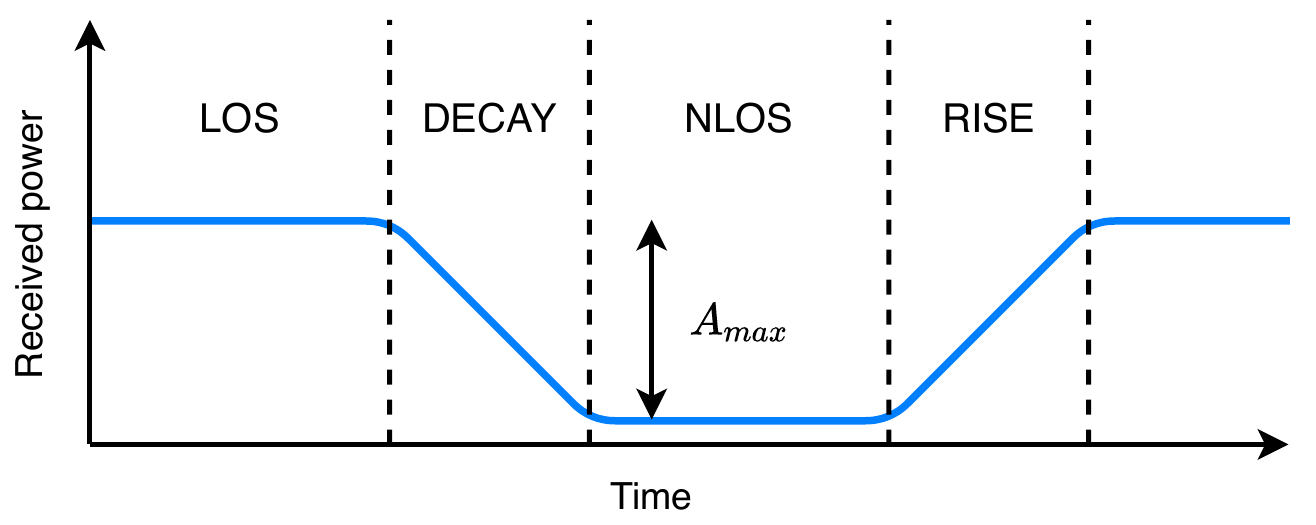}
	\caption{Four-state blockage model.}
	\label{fig:four_state_blockage_model}
\end{figure}

The \ac{LOS} path between \ac{AP} and \ac{UE} can be obstructed by blockage, such as a moving pedestrian or the user body itself, which can decrease the received signal power.
We assume the blockage event is modelled according to the four-state model described in \cite{maccartney2017rapid} and affects the channel quality of the whole link bandwidth for each UE independently.
We denote as $s\in \{ \chst{LOS},\chst{NLOS},\chst{DECAY},\chst{RISE}\}$ the states of the \ac{AP}-\ac{UE} link and we represent them in Fig. \ref{fig:four_state_blockage_model}.
The \chst{LOS} state denotes the link having \ac{LOS} propagation conditions, 
while the \chst{NLOS} state denotes the link being obstructed by blockage.
Moreover, the \chst{DECAY} state denotes the transition state from \chst{LOS} to \chst{NLOS} and is characterised by a decay factor \unit[$\rho_\mathrm{D}$]{dB/s}.
The \chst{RISE} state denotes the transition state from \chst{NLOS} to \chst{LOS} and is characterised by a factor \unit[$\rho_\mathrm{R}$]{dB/s}.

We assume that the blockers' arrivals follow a Poisson process with average arrival rate \unit[$\lambda_\mathrm{B}$]{blockers/s}, 
and the blockage state has duration which is exponentially distributed with average value \unit[$\tau_\mathrm{B}$]{ms} \cite{jain2019impact}. 
The blockage attenuation power $A(t)$ varies according to the channel state and can be expressed as
\begin{equation}
A(t) = 
\begin{dcases}
0 & \text{if $s$=\chst{LOS},} \\
\rho_\mathrm{D} \cdot (t-t_\mathrm{D}) & \text{if $s$=\chst{DECAY},} \\
A_\textrm{max} & \text{if $s$=\chst{NLOS},} \\
A_\textrm{max}-\rho_\mathrm{R} \cdot (t-t_\mathrm{R}) & \text{if $s$=\chst{RISE},}
\end{dcases}
\label{eq:blockageAttenuation}
\end{equation}
where $t_\mathrm{D}$ and $t_\mathrm{R}$ represent the time instants when the states \chst{DECAY} and \chst{RISE} start.
We also assume that if multiple blockers arrive together, the channel remains in \chst{NLOS} state for the duration of the longest blockage event.



The link budget for the channel between AP and \ac{UE} $u$ in the time-slot $t$ can expressed as,
\begin{equation}
    p_\textrm{rx,u}(t) = p_\textrm{tx} \cdot g_\textrm{tx} \cdot g_\textrm{rx} \cdot\ell_0 \cdot d_u^{-\nu} \cdot A(t),
\end{equation}
where $p_\textrm{tx}$ is the transmit power, $g_\textrm{tx}$ is the transmit antenna gain, $g_\textrm{rx}$ is the receive antenna gain, $\ell_0$ is the path loss at one metre distance under free space propagation conditions, $d_u$ is the \ac{AP}-\ac{UE} distance and $\nu$ is the attenuation exponent.

We define the signal-to-noise ratio (SNR) as:
\begin{equation}
z_{u}(t) = \frac{p_\textrm{rx,u}(t)}{p_\textrm{n}},
\label{eq:snr}
\end{equation}
where $p_\textrm{n}$ represents the noise power over bandwidth $b$.
The instantaneous user data rate can be expressed as:
\begin{equation}
    r_u(t) = b \cdot \log_2(1+z_{u}(t)) \cdot x_u(t) \cdot \Ib(z_{u}(t)>\zeta),
    \label{eq:insta_user_rate}
\end{equation}
where $\Ib(\cdot)$ is a binary function that indicates when the link goes to outage, i.e. $z_{u}(t)<\zeta$, and the data rate becomes $r_u(t)=0$.

\subsection{Conventional Scheduling Algorithms}

The conventional \ac{PF} scheduler aims to provide a fair distribution of resources among the set of UEs that are served by the \ac{AP}.
Within the pool of UEs to schedule there can be \acp{UE} experiencing favourable channel conditions and \acp{UE} with less favourable channel conditions. 
To obtain a fair allocation, the \ac{PF} scheduler gives priority to the \acp{UE} that have the highest ratio between the feasible rate at the current time-slot and the average rate over former successive time-slots \cite{kim2005proportional,liu2008proportional}.
The \ac{PF} scheduler policy to assign the user $u^*$ in the time-slot $t$ can be expressed as
\begin{equation}
    u^* = \operatorname*{arg\,max}_{u} \frac{r_u(t)}{\bar{r}_u(t-1)},
    \label{eq:userpriority_pf}
\end{equation}
where $\bar{r}_u(t)$ is the average rate of user $u$ in the time slot $t$, and is calculated based on an exponential moving average:
\begin{equation}
\begin{split}
    \bar{r}_u(t) &= (1-w)\,\bar{r}_u(t-1) + w\,r_u(t),
\end{split}
\end{equation}
where $w \in [0,1]$ is a system parameter that weights the importance of the current feasible rate with respect to the average rate when computing the average rate metric.

On the other hand, the \ac{MaxMin} scheduler aims to maximise the minimum rate of the \acp{UE}.
This is done by prioritising the \ac{UE} with the lowest average rate.
The \ac{MaxMin} scheduler policy to assign the user $u^*$ in the time-slot $t$ can be expressed as
\begin{equation}
    u^* = \operatorname*{arg\,max}_{u} \frac{1}{\bar{r}_u(t-1)}.
    \label{eq:userpriority_mm}
\end{equation}

Under the effect of blockage, the policies described in Equations (\ref{eq:userpriority_pf}) and (\ref{eq:userpriority_mm}) may lead to a decrease of the UE average rate. 
This effect results from the detrimental impact of the blockage attenuation on the feasible data rate. 
In detail, the above scheduling policies' specific behaviours can be described as follows:
\begin{itemize}
    \item The \ac{PF} priority function in (\ref{eq:userpriority_pf}) is directly proportional to the current feasible rate.
    Thus, as the feasible rate decreases with blockage attenuation, the priority of \acp{UE} suffering from blockage decreases, giving more priority to \acp{UE} in favourable conditions.
    This decrease in the priority is not immediate since the priority is also inversely proportional to the average rate that decreases slower than the feasible rate, depending on the weight of the moving average.
    \item The \ac{MaxMin} priority function in (\ref{eq:userpriority_mm}) is inversely proportional to the average rate.
    It means that, even when the \ac{UE} is in outage, the priority keeps increasing from slot to slot as the average rate decreases.
    Thus, more slots are allocated to the \ac{UE} suffering from blockage and fewer slots to \acp{UE} in favourable conditions.
\end{itemize}
Therefore, both the \ac{PF} and \ac{MaxMin} schedulers suffer in case of blockage. In the first case, only a few slots are allocated by the \ac{PF} scheduler when it is most needed, while in the second case too many slots are allocated by the \ac{MaxMin} scheduler when they are not needed.

\subsection{Blockage-Aware Proportional Fair Scheduling in \ac{mmWave} Spectrum}

To circumvent the scheduling inefficiencies caused by the blockage effect, we study the use of a predictive scheduler that leverages the estimation of the channel status in future slots similarly to the ones proposed in \cite{bang2008channel,hajipour2010proportional,schmidt2011prediction,shen2019proactive}.
Differently from \cite{bang2008channel,hajipour2010proportional,schmidt2011prediction,shen2019proactive}, 
we study the application to the \ac{mmWave} network use case, where it is potentially possible to predict the received power degradation due to temporary link blockage \cite{nishio2019proactive}.
Therefore, we assume that the system can predict the received power $n_T$ slots in advance, where $n_T$ is the prediction window.
We model the potential error due to inaccurate predictions with a random variable $E\sim\mathcal{N}(0,\sigma^{2})$ 
\cite{hajipour2010proportional,duel-hallen2006long} 
which is added to the future ground-truth values of the received power obtained from simulations to model the predicted received power $\hat{p}_\textrm{rx,u}(t)$. 
We also assume that $\sigma$ increases with the prediction window size given that the prediction accuracy decreases \cite{nishio2019proactive}.

Hence, the estimated \ac{SNR} $\hat{z}_{u}(t')$ at a future time-slot $t' < t+n_T$ can be expressed as:
\begin{equation}
    \hat{z}_{u}(t') = \frac{p_\textrm{rx,u}(t') + E}{p_\textrm{n}}.
    \label{eq:est_snr}
\end{equation}
We assume that a blockage is detected by the predictor when the estimated received power $\hat{p}_\textrm{rx,u}(t')$ shows significant variations within the prediction window. Conversely, we assume that there is no blockage when $\hat{p}_\textrm{rx,u}(t')$ remains constant over time.\footnote{As the main focus of our work is to evaluate the scheduling performance under blockage effect, we assume that in this type of scenario the attenuation due to the blockers over time prevails on other time-dependent components such as fast fading or interference.}

Once a blockage is predicted, and given the estimated channel quality, a predictive scheduler will allocate more resources to those \acp{UE} which are to suffer from the blockage.
In practice, the \ac{UE} priority is modified expressing the denominator of the \ac{BA-PF} scheduler policy as a sum of the predicted feasible rates at future time-slots.
For the future time-slots at the end of the prediction window, the predicted rate for the blocked \acp{UE} is expected to be low due to the blockage attenuation, making the denominator smaller than it would have been for if the \acp{UE} had not been affected by the blockage. 
Consequently, the \acp{UE} predicted as blocked are assigned a higher priority than the other \acp{UE} and are likely to be granted more scheduling opportunities for the future time-slots at the beginning of the prediction window. 
For those time-slots, their channels still experience favourable propagation conditions.
On the other hand, the \acp{UE} which are predicted as non-blocked may have more scheduling opportunities in the last slots of the prediction window, when the channels of the blocked \ac{UE} will be affected by a strong blockage attenuation and are assigned a low priority which allows to release the resources to those non-blocked \acp{UE}.

Hence, we adapt the priority function in (\ref{eq:userpriority_pf}) to make the \ac{PF} scheduler blockage-aware (\ac{BA-PF}).
The \ac{BA-PF} scheduler policy to assign the user $u^*$ in the time-slot $t'$ can be expressed as
\begin{equation}
    u^* = \operatorname*{arg\,max}_{u} \frac{\hat{r}_u(t')}{S_u(t')},
    \label{eq:userprioirty_ba}
\end{equation}
where $\hat{r}_u(t')$ is the feasible instantaneous user rate at a future time-slot $t' \in [t+1, t+n_T]$ and can be computed as:
\begin{equation}
    \hat{r}_u(t') = b \cdot \log_2(1+\hat{z}_{u}(t')) \cdot \Ib(\hat{z}_{u}(t')>\zeta).
    \label{eq:feasible_insta_user_rate}
\end{equation}
and $S_u(t')$ is the estimated sum-rate of all time-slots from $t'$ to $t+n_\textrm{T}$ given the allocation vector $\mathbf{X}_u$ and can be expressed as:
\begin{align}
    S_u(t') = \sum\limits_{j=t'}^{t+n_\textrm{T}} \hat{r}_u(j) \cdot x_u(j).
\end{align}
Thus, the assignment of the first slots depends on how the following slots in the future are assigned.
Because of that, the time-slot assignment starts from the end of the prediction window at time-slot $t'=t+n_\textrm{T}$ and works backwards to time-slot $t'=t+1$, as shown in Algorithm \ref{alg:scheduler}.
\begin{algorithm}
\SetAlgoLined
\KwResult{Allocation assignment $x$}
 $x_u(t') = 0, \quad \forall \; u,t'$\;
 \eIf{blockage is predicted to happen during any of the next $n_T$ slots}
 {
 \For{$t'=t+n_T$ down to t}{
        \For{$u=1$ to $n_U$}{
            $S_u(t') = \sum\limits_{j=t'}^{t+n_\textrm{T}} \hat{r}_u(j) \cdot x_u(j)$\;
        }
    $u^* = \operatorname*{arg\,max}_{u} \frac{r_u(t')}{S_u(t')}$\;
    $x_{u*}(t')=1$\;
 }
 }
 {Execute \ac{PF} algorithm.}
 \caption{\ac{BA-PF} scheduling algorithm}
 \label{alg:scheduler}
\end{algorithm}

For instance, consider a \ac{UE} $u$ that suffers from blockage and has the estimated rate ($\hat{r}_u$) degraded in the future.
Because of the backward iteration, other \acp{UE} were allocated in the last slots as $\hat{r}_u$ is lower than the estimated rates of the other \acp{UE}.
Hence, as the iteration reaches the first slots, $\hat{r}_u$ is high and $S_u$ is low since very few time-slots have been allocated to \ac{UE} $u$, leading to a increased priority function according to (\ref{eq:userprioirty_ba}).
Therefore \ac{UE} $u$ has the transmission anticipated as it gets more priority in the first slots and releases the last slots for \acp{UE} with better channel conditions. 

\section{Numerical Results}
\label{sec:numericalresults}
\subsection{Simulation Setup}

\begin{figure}[h]
	\centering
	\includegraphics[width=.7\linewidth]{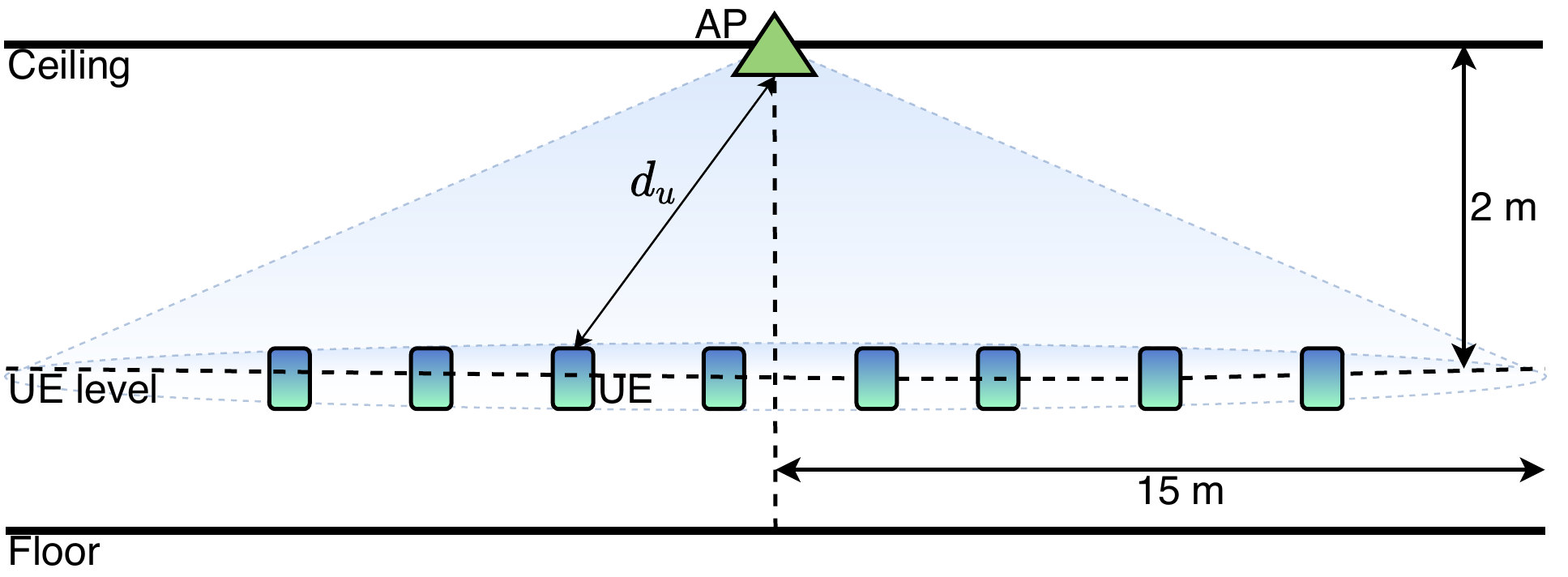}
	\caption{Ceiling-mounted \ac{AP} setup with downward facing beam illuminating the \acp{UE}.}
	\label{fig:ceiling_mounted_blockage}
\end{figure}

In this section, we evaluate and compare the \ac{PF}, \ac{MaxMin} and \ac{BA-PF} schedulers performance.
As depicted in Fig. \ref{fig:ceiling_mounted_blockage}, we consider a single ceiling-mounted \ac{AP} that is positioned in the centre of the network area at a height of \unit[2]{m} above the \ac{UE} antennas and is equipped with a directive antenna facing downward having beamwidth \unit[170]{$^\circ$} and gain of \unit[3.16]{dBi} \cite{firyaguna2020performance}.
The \acp{UE} are uniformly distributed in the \ac{AP} coverage area (of radius \unit[$\rho=$15]{m}), are equipped with an omnidirectional receiver antenna and are all served from the same \ac{AP}.
The channel model parameters are in line with empirically-tested model described in \cite{yoo2017measurements} having \ac{LOS} channel state with parameters \unit[$\ell_0=$ 63.4]{dB} and \unit[$\nu=$ 1.72].
We assume maximum blockage attenuation \unit[$A_\mathrm{max}=40$]{dB}, decay rate \unit[$\rho_\mathrm{D}=0.2$]{dB/ms}, and rise rate \unit[$\rho_\mathrm{R}=6.7$]{dB/ms}, in line with the measurements reported in \cite{ju2019millimeter}.

We simulate a sequence of 48000 time-slots of an \ac{OFDMA} frame transmission with carrier frequency of \unit[60]{GHz} according to the 5G \ac{NR} standard \cite{3gpp_tr_38211}.
We gain insights on the various schedulers by changing the characteristics of the blockage scenario, including different blockers' arrival rate $\lambda_\mathrm{B}$ and blockage duration $\tau_\mathrm{B}$.
To analyse the impact of the blockage effects on scheduling, we evaluate the empirical cumulative distribution function (ECDF) of the average user data rate.
The 1st percentile rate, mean rate and fairness metrics are selected for performance evaluation.
The 1st percentile rate can be interpreted as the average data rate of the \acp{UE} that are most affected by blockage.
The mean rate is defined as the sum of the average user rates divided by the number of \acp{UE} in the cell,
and fairness is defined as the Jain's fairness index \cite{jain1984quantitative}.
Settings for the system parameters are provided in Table \ref{tab:systemparameters_ch4}.

\begin{table}[ht!]
\caption{System parameters}
\label{tab:systemparameters_ch4}
\centering
\begin{tabular}{llr}
\toprule
Number of \acp{UE} & $n_U$ & 8 \\
Slot duration & $\Delta t$ & \unit[62.5]{$\mu$s} \\
Frame bandwidth & $b$ & \unit[2]{GHz} \\
Moving average weight & $w$ & 0.5 \\
Transmit power & $p_\textrm{tx}$ & \unit[100]{mW} \\
Noise figure & & \unit[9]{dB} \\
Noise power & $p_n$ & \unit[-71.99]{dBm} \\
SNR threshold & $\zeta$ & \unit[0]{dB} \\
\ac{AP} beamwidth &  & \unit[170]{$^\circ$} \\
\ac{AP} antenna gain & $g_{tx}$ & \unit[3.16]{dBi} \\
\ac{UE} antenna gain & $g_{rx}$ & \unit[0]{dBi} \\
Prediction window & $n_T$ & \unit[\{50, 200, 500\}]{ms} \\
Prediction error & $\sigma$ & \{ $10^{-3}$,$10^{-2}$,$10^{-1}$\} \\
\bottomrule
\end{tabular}
\end{table}

\subsection{User Performance}
\begin{figure}
\centering
\begin{subfigure}{.85\linewidth}
    \centering
    \includegraphics[width=\linewidth,trim={0cm 0cm 0cm 0cm},clip]{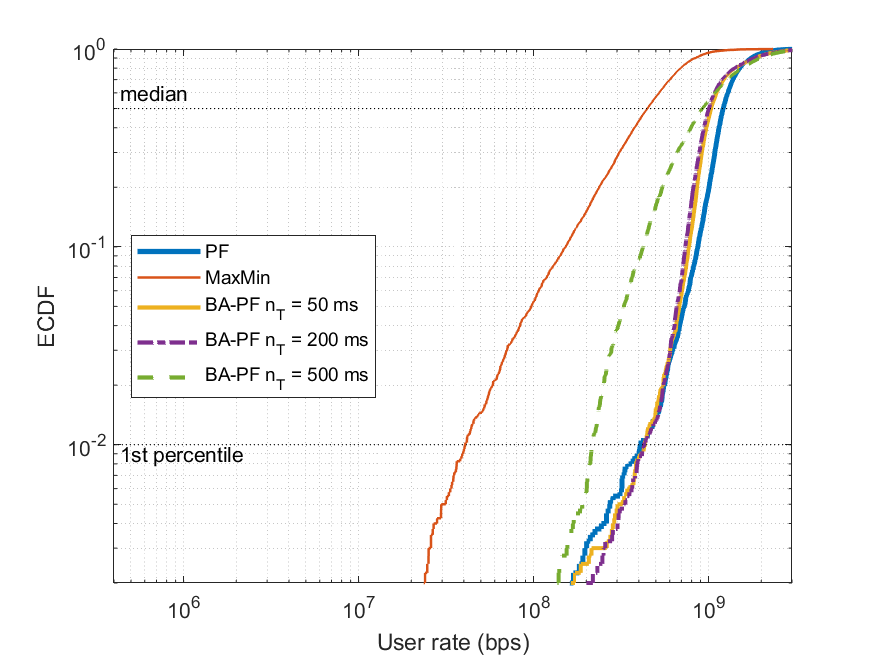}
    \caption{\unit[$\lambda_\mathrm{B}=0.2$]{blockers/s}, \unit[$\tau_\mathrm{B}=1000$]{ms}}
    \label{fig:ecdf_a-1_d-1}
\end{subfigure}
\hfill
\begin{subfigure}{.85\linewidth}
    \centering
    \includegraphics[width=\linewidth,trim={0cm 0cm 0cm 0cm},clip]{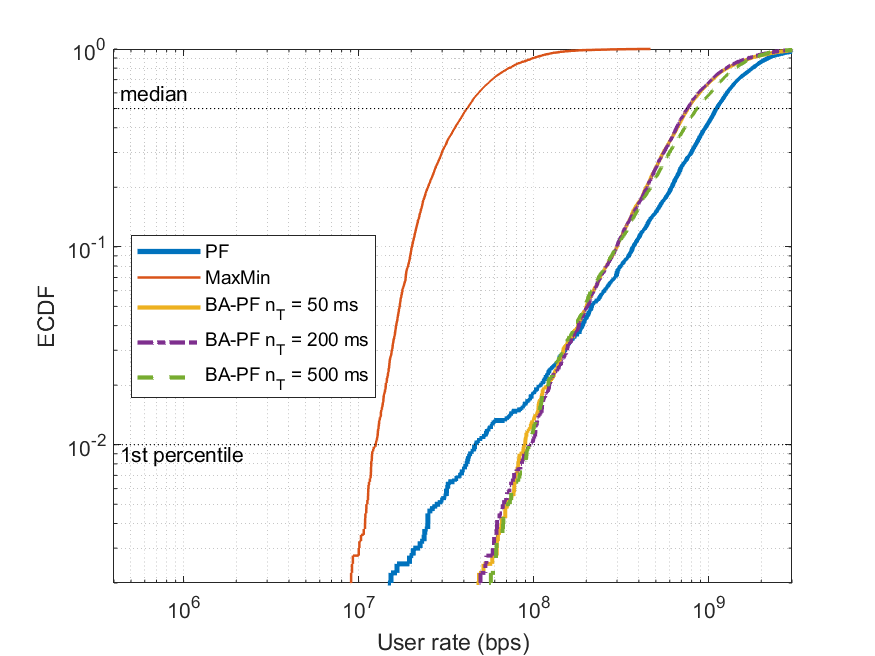}
    \caption{\unit[$\lambda_\mathrm{B}=1.0$]{blockers/s}, \unit[$\tau_\mathrm{B}=1000$]{ms}}
    \label{fig:ecdf_a-2_d-1}
\end{subfigure}
\hfill
\begin{subfigure}{.85\linewidth}
    \centering
    \includegraphics[width=\linewidth,trim={0cm 0cm 0cm 0cm},clip]{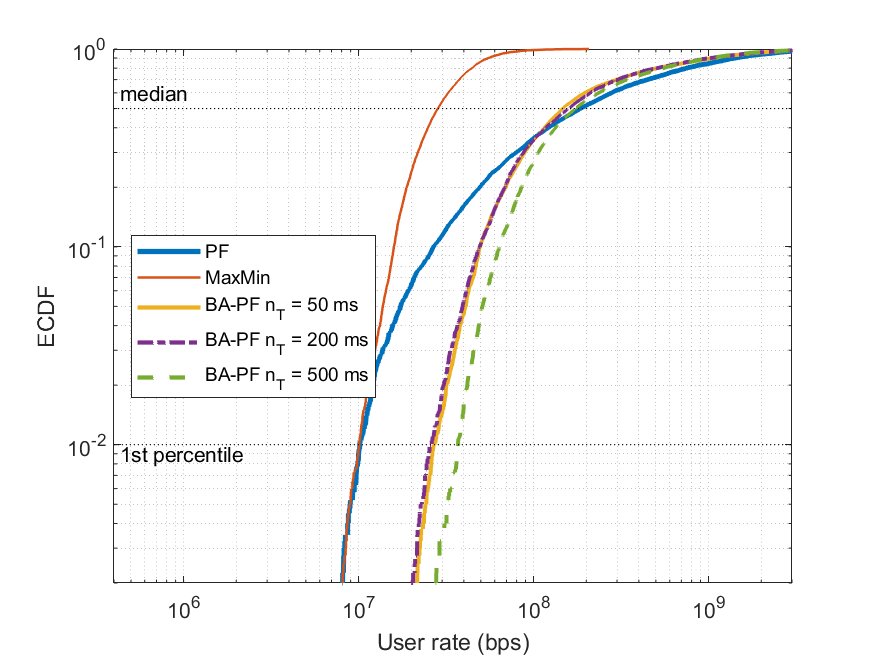}
    \caption{\unit[$\lambda_\mathrm{B}=2.0$]{blockers/s}, \unit[$\tau_\mathrm{B}=3000$]{ms}}
    \label{fig:ecdf_a-3_d-2}
\end{subfigure}
    \caption{Empirical cumulative distribution function of the user rate for different blockage scenarios with eight \acp{UE} in the cell.}
    \label{fig:ecdf_rate}
\end{figure}

In Fig. \ref{fig:ecdf_rate} we show the ECDFs of the average user rate for the \ac{PF}, \ac{MaxMin} and \ac{BA-PF} schedulers in three different blockage scenarios: 
(a) low blockers' arrival rate, short blockage duration (Fig. \ref{fig:ecdf_a-1_d-1});
(b) mid-range blockers' arrival rate, short blockage duration (Fig. \ref{fig:ecdf_a-2_d-1}); and
(c) high blockers' arrival rate, long blockage duration (Fig. \ref{fig:ecdf_a-3_d-2}).

In the blockage scenario (a) depicted in Fig. \ref{fig:ecdf_a-1_d-1},
the \ac{MaxMin} scheduler (red curve) has the worst performance with respect to \ac{PF} and \ac{BA-PF} schedulers.
This is because the \acp{UE} that suffered from blockage were mostly prioritised even though they had poor channel quality, leading to low user rate. On the contrary, the \acp{UE} with better channel quality were less prioritised, also leading to low user rate.
Moreover, we observe a smaller discrepancy in performance between \ac{PF} (blue curve) and \ac{BA-PF} (yellow, purple and green curves) schedulers.
In fact, when there are fewer blockage events, this is expected since the \ac{BA-PF} priority function in (\ref{eq:userprioirty_ba}) is executed only a few times.
In the vast majority of cases, the conventional \ac{PF} is executed during the prediction window because no blockage is detected.

Furthermore, in the blockage scenario (b) depicted Fig. \ref{fig:ecdf_a-2_d-1}, where the blockage effect is more severe, the \ac{MaxMin} scheduler (red curve) still has the worst performance, while the difference between \ac{PF} and \ac{BA-PF} performances increases when looking at the 1st percentile of the ECDF, where we observe a performance gain of \unit[48]{\%} with respect to the \ac{PF} for all the prediction window values.
This is because the \ac{BA-PF} priority metric in (\ref{eq:userprioirty_ba}) is designed to allocate resources at the beginning of the prediction window and to release them at the end considering that there is only one transition from \chst{LOS} to \chst{NLOS} states.
Hence, when the blockage duration is short, or there are not as many arrivals as in the blockage scenario (c), multiple state transitions may occur within a large prediction window, leading to a non-ideal scenario for the \ac{BA-PF} scheduling metric as the resource anticipation is not effective.

Finally, the best performance can be observed for the blockage scenario (c), as depicted in Fig. \ref{fig:ecdf_a-3_d-2}, when it is more likely that the blockage state is long enough to occur only once within the prediction window.
When using the \ac{PF} scheduler, the \acp{UE} that are affected by the blockage are less prioritised as more blockage events occur, decreasing their user rate. 
Thus, the performance of the \ac{PF} scheduler (blue curve) approaches the performance of the performance of the \ac{MaxMin} scheduler (red curve) for the lower values of the ECDF.
In contrast, when adopting the \ac{BA-PF} scheduler, the \acp{UE} that suffer from blockage are more prioritised than in the \ac{PF} scheduler, and we observe gains up to \unit[153]{\%} in the 1st percentile rate with respect to the \ac{PF} performance for the prediction window value \unit[$n_T=50$]{ms} (yellow curves).
Moreover, this gain is improved to \unit[267]{\%} when increasing the prediction window to \unit[$n_T=500$]{ms} (green curve in Fig. \ref{fig:ecdf_a-3_d-2}), as the pool of resources to allocate increase and there is more time to anticipate a transmission for a blocked \ac{UE}.

\subsection{Network Performance}

\begin{figure}
	\centering
	\includegraphics[width=\linewidth]{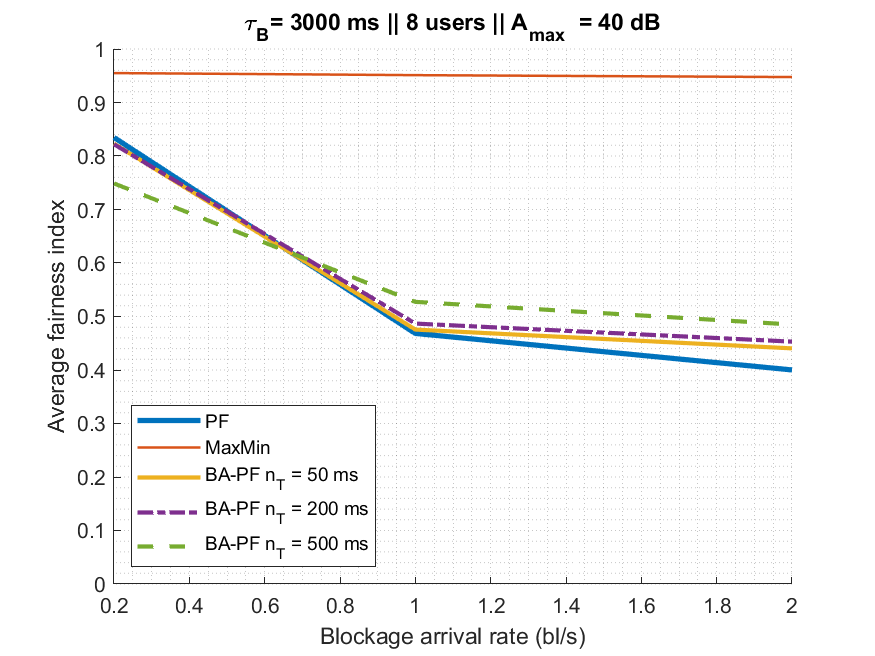}
	\caption{Average fairness performance as a function of $\lambda_\mathrm{B}$, with \unit[$\tau_\mathrm{B}=3000$]{ms}.}
	\label{fig:avg_fair}
\end{figure}
\begin{figure}
	\centering
	\includegraphics[width=\linewidth]{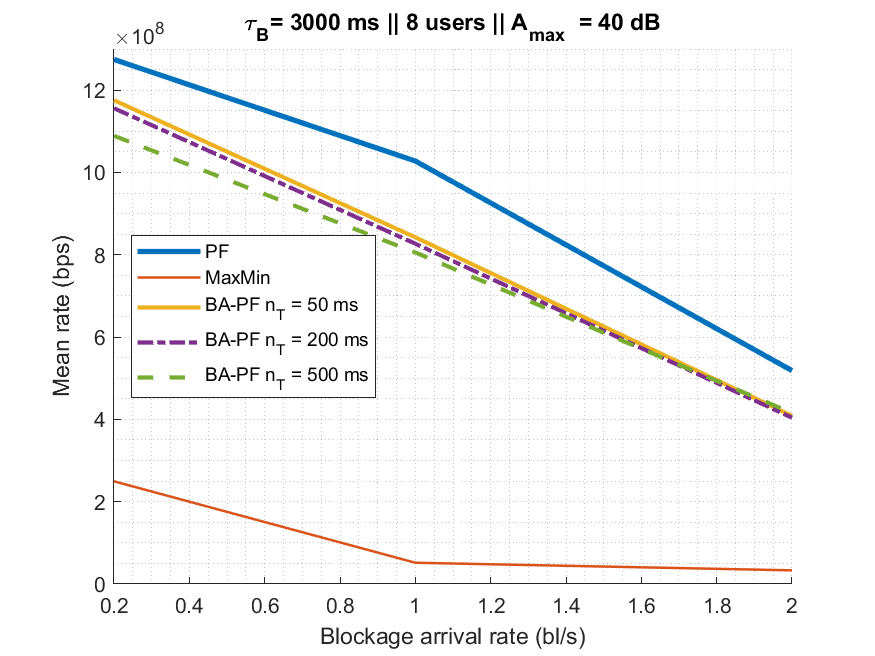}
	\caption{Mean rate performance as a function of $\lambda_\mathrm{B}$, with \unit[$\tau_\mathrm{B}=3000$]{ms}.}
	\label{fig:mean_rate}
\end{figure}

In Figs. \ref{fig:avg_fair} and \ref{fig:mean_rate}, we show the performance of the network in terms of mean rate and fairness index as a function of the blockage effect.
From Fig. \ref{fig:avg_fair} we can observe that the \ac{MaxMin} fairness performance is high (0.95) and is not affected by blockage. 
The \ac{BA-PF} performance degrades with $\lambda_\mathrm{B}$, as the blockage scenario gets more severe.
This is because the preemptive allocation for \acp{UE} suffering from blockage drains a significant amount of resources taken from the pool of all the available resources.
Hence, as the blockage severity increases, \ac{BA-PF} assigns more resources to \acp{UE} transitioning to blockage state, disfavouring other \acp{UE}, and consequently decreasing fairness.
However, the \ac{BA-PF} fairness is still higher than \ac{PF} when the blockage arrival rate is high (\unit[$\lambda_\mathrm{B} \geq 1$]{blockers/s}), as \ac{PF} prioritises the allocation of resources to a decreasing number of \acp{UE} that are not blocked.

Moreover, from Fig. \ref{fig:mean_rate}, we can observe that the \ac{MaxMin} mean rate is significantly lower compared to other schedulers, as the high fairness is achieved by taking away resources from \acp{UE} with good channel quality and giving them to \acp{UE} with worse quality, levelling down the user rates. 
We observe that \ac{BA-PF} and \ac{PF} mean rates significantly degrade with $\lambda_\mathrm{B}$, as the number of blocked \acp{UE} increases.
We see that the \ac{BA-PF} mean rate is up to \unit[22]{\%} lower than the \ac{PF} mean rate, as the gains achieved by improving the data rate of \acp{UE} suffering from blockage comes at the cost of reducing the data rate of non-blocked \acp{UE}.

\section{Conclusion}
\label{sec:conclusion}
In this work, we assessed the behaviour of schedulers in the event of a blockage affecting the \ac{LOS} signal path.
We showed that the conventional \ac{MaxMin} scheduler, designed to improve the cell-edge user rate, cannot improve the rate of the users suffering from blockage and has its performance significantly degraded.
We showed that the \ac{PF} scheduler, although it is less affected by blockage compared to \ac{MaxMin}, can have a user performance as low as \ac{MaxMin} when blockage scenario is severe.
This is because conventional schedulers react to blockage by allocating more resources to compensate the loss in the data rate (as the \ac{MaxMin} scheduler does), or by allocating fewer resources to prioritise other \acp{UE} with better channel conditions (as the \ac{PF} scheduler does).
However, when the \ac{UE} is entirely in outage, a successful transmission might be impossible to achieve, and any amount of allocated resources may be unused, reducing the efficiency of the allocation.

This reduction in efficiency suggests that conventional schedulers should be adapted to mitigate the blockage effects in the context of 5G-mmWave networks.
We therefore considered an adapted scheduling metric \ac{BA-PF} which relies on blockage prediction to compute the estimated user rates, many slots ahead of a blockage event.
The effectiveness of this adaptation is demonstrated by increasing the blockage arrival rate and the blockage duration.
We showed that the gain in performance between the \ac{BA-PF} scheduler and the conventional ones is more significant when the arrival rate and duration are high, i.e., when the channel stays in blockage state long enough to allow for an effective anticipation of resources.
This gain is due to an improvement in the rate of users affected by blockage, which is reflected in the gains in the 1st percentile rate and fairness with respect to the \ac{PF} performance.
However, this improvement comes at the expense of performance experienced by other \acp{UE}, as reflected on the losses in mean rate performance.




\ifCLASSOPTIONcaptionsoff
  \newpage
\fi



\bibliographystyle{IEEEtran}
\bibliography{myreferences}
\end{document}